# Creating Tailored and Adaptive Network Services with the Open Orchestration C-RAN Framework

Marti Floriach-Pigem, Guillem Xercavins-Torregrosa,
and Antoni Gelonch-Bosch
Dept. Signal Theory and Communications
Barcelona Tech-UPC, Castelldefels, Spain
{mfloriach90, guillemxercavins}@gmail.com,
antoni@tsc.upc.edu

Vuk Marojevic
Wireless@Virginia Tech
Bradley Dept. Electrical and Computer Engineering
Virginia Tech, Blacksburg, VA, USA
maroje@vt.edu

*Abstract*—Next generation wireless communications networks will leverage software-defined radio and networking technologies, combined with cloud and fog computing. A pool of resources can then be dynamically allocated to create personalized network services (NSs). The enabling technologies are abstraction, virtualization and consolidation of resources, automatization of processes, and programmatic provisioning and orchestration. ETSI's network functions virtualization (NFV) management and orchestration (MANO) framework provides the architecture and specifications of the management layers. We introduce OOCRAN, an open-source software framework and testbed that extends existing NFV management solutions by incorporating the radio communications layers. This paper presents OOCRAN and illustrates how it monitors and manages the pool of resources for creating tailored NSs. OOCRAN can automate NS reconfiguration, but also facilitates user control. We demonstrate the dynamic deployment of cellular NSs and discuss the challenges of dynamically creating and managing tailored NSs on shared infrastructure.

*Keywords*—Management and orchestration, network functions virtualization, software-defined radio, testbed.

## I. INTRODUCTION

Cloud computing has experienced enormous growth and has been widely deployed to provide differentiated and virtually unlimited computing services to companies and individuals alike. Along with the Cloud, new service models were introduced: infrastructure, platform and software as a service (IaaS, PaaS and SaaS). These models provide capital and operational expenditure (CAPEX and OPEX) benefits and economy of scale through infrastructure sharing and "pay-per-use" business models.

Wireless communications networks have started to adopt these principles since 2010 [2]-[4], taking benefit of software-defined radio (SDR) technology to address the challenging implementation at the physical layer. The Cloud radio access network or C-RAN adopts the IaaS model to centralize the baseband processing of multiple base stations in a shared data center. It offers an attractive alternative to service providers because of cost reductions and flexibility of operation. Whereas cloud technology is mature and effectively provides computing, storage and application services for a wide set of use cases, it is not yet meeting the requirements of legacy and next generation RANs and communications services. Rather, the cellular communications industry limits the use of C-RANs to centralized baseband processing.

A wireless network service (NS) refers to a network that is built ad hoc to deliver the desired user service running on top of it. A NS that provides the voice service, for example, would look different than a NS that provides video streaming. The desired user services should drive the selection or creation of tailored NSs to optimize networks and resources.

The physical resources needed for provisioning a NS are wireless infrastructure, radio frequency (RF) spectrum, transmission power, and so forth. Today, most of these resources are proprietary and heavily regulated. Doyle et al. [5] argue for the convenience of providing wireless services on demand. A pool of physical resources, which include antenna sites, networking components, and processing nodes, are offered to mobile virtual network operators (MVNOs) to deploy their virtual networks that are most suited for the services they intended to deliver. Ideally, the most-suitable infrastructure needed for creating the desired NS is served just in time to the wireless operator.

The role of MVNOs will change to a more dynamic operator. In this new role, MVNOs will access software images of NSs, describing the necessary wireless infrastructure, among others, and dynamically customize and adapt the NS to the changing operational conditions: traffic, service demand, channel conditions, coverage, etc. Such adaptation implies changes in the wireless infrastructure that is leased from the wireless infrastructure provider (WIP). This approach is compliant with cloud computing, but adds management challenges to meet the strict quality of service (QoS) requirements of communications services. Some of these challenges are related to RF isolation, scalability and real-time wireless network reconfiguration.

The European Telecommunications Standards Institute (ETSI) is making important efforts to facilitate managing C-RANs [6]. ETSI's network functions virtualization (NFV) management and orchestration (MANO) framework provides the architecture (Fig. 1) and specifications of such a management layer. The structure that ETSI defines is composed of three main blocks: the orchestrator, the virtual network function (VNF) manager and the virtualized infrastructure manager (VIM). The orchestrator is the high-level management entity that is responsible for managing the overall network. The VNF manager oversees the lifecycle management of VNF instances, coordinating and adapting them to the NFV infrastructure (NFVI). The VIM controls and manages the NFVI's physical computing, storage and networking resources (Fig. 1).

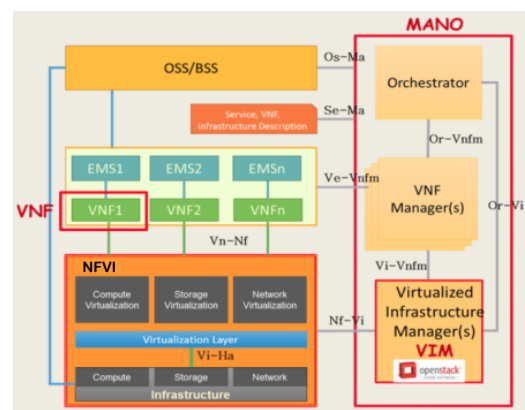

Fig. 1. ETSI MANO NFV framework.

NFV is a major R&D topic today that received industry attention early on. Intel [7] describes the implementation of a virtual machine (VM) executing the evolved packet core (EPC), the core network (CN) of 4G long-term evolution (LTE), and the 3G CN using OpenStack as the VIM and the Kernel-based VM (KVM) as the hypervisor. The design, implementation and evaluation of two EPC architectures, one based on the principles of software-defined networking (SDN) and the other based on NFV, are presented in [8]. The virtualization process of a base station and a CN is discussed in [9]. Bouras et al. [10] compare SDN and NFV based solutions for mobile radio environments.

Next generation wireless networks are going to adopt virtualization technology to enable infrastructure sharing. This paper considers the new role for MVNOs, which will rent all resources (spectrum, antenna sites, computing capacity, connectivity, etc.) from the resource pools to dynamically create and deploy the NS that best satisfies the end user service demands at a given place, time, and as a function of the environmental conditions [5].

The existing frameworks, such as Open Source MANO [11], are rather heavy and require a certain infrastructure. We designed and developed the Open Orchestration C-RAN (OOCRAN) framework [12] as a lightweight alternative that allows building small-scale testbeds using different virtualization tools to lower the barrier of entry. OOCRAN extends the resource management functionalities of existing MANO NFV contributions by incorporating radio resources into the infrastructure resource pool.

## II. THE OOCRAN FRAMEWORK

OOCRAN is an orchestration layer that follows the ETSI MANO architecture and extends its functionalities for creating, coordinating and managing wireless networks. An OOCRAN user can take the role of a WIP and dynamically provision the desired wireless infrastructure, software and hardware, to the wireless service provider (WSP) in order to create the NS tailored to its user demands. This framework enables analyzing different ways of splitting the resource deployment, management and maintenance responsibilities and evaluating different shared resource access policies. OOCRAN can easily create a complete communications system or NS by chaining several VNFs and taking into account the available physical resources, including RF equipment, networking devices and computing systems.

In accordance with the ETSI MANO architecture, the OOCRAN framework has been developed using OpenStack, Newton release as the VIM, and KVM, a high performance open source platform, as the hypervisor. Through a user interface based on Web services, a user can activate monitoring tools, schedulers, queues, alarms and scripting commands to properly and dynamically manage VNFs and their configurations. Using these tools, a user can build a detailed control layer of the virtual infrastructure that supports a specific NS. Different management policies can be defined to specify actions as a function of the current state of the NS or the physical wireless infrastructure. All source code is released under the AGPL license and is available for free download from [12].

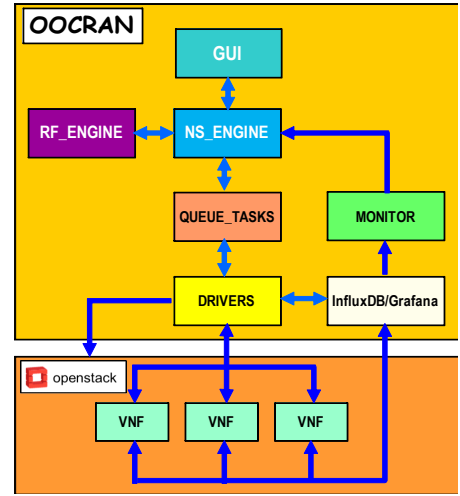

Fig. 2.  OOCRAN software architecture and components.

### A. OOCRAN Software Management Framework

The OOCRAN software management framework is built around six main components: GUI, MONITOR, NS_ENGINE, RF_ENGINE, QUEUE_TASKS and DRIVERS. Their functionalities and interactions have been defined to facilitate the design and application of optimized management algorithms for the creation and deployment of tailored NSs. Fig. 2 illustrates the OOCRAN software architecture and a brief description of components and functionalities follows:

- **GUI** provides a user-friendly environment for a human operator to manage the OOCRAN actions. Developed using Python 2.7 and Django, it allows saving the NSs and VNFs and defining actions and tasks to be executed when changing the NS.
- **MONITOR** interfaces with third-party programs Grafana and InfluxdB to capture the alarms they activate about the state of VNFs or NSs. It identifies the alarms, checks the credentials and sends feedback to the NS_ENGINE, which decides on the corresponding action.
- **NS_ENGINE** manages the entire OOCRAN framework and actuators and applies its actions to NSs and VNFs. Its functionalities include the definition of the NS/VNF lifecycle and the actions when the NS reaches a specific state or when an alarm is activated.
- **RF ENGINE** manages the connections between the VNF and the RF frontends. It installs the necessary third party libraries in the VNFs to create those connections. It manages the pool of RF resources and handles frequency reuse, among others.
- **QUEUE_TASK** is developed using a task manager called RabbitMQ and processes command queues allowing to execute OOCRAN tasks asynchronously. It supports the NS_ENGINE functionalities by hosting the command queues that NS_ENGINE generates and, when scheduled, delivers the command to the DRIVERS module.
- **DRIVERS** uses third-party APIs (Graphana and InfluxDB) to execute the NS_ENGINE commands.

OOCRAN uses Vagrant OpenStack compatible VIMs, such as Virtualbox, KVM, Docker, or VMware. Fig. 2 shows the interfaces among OOCRAN components, where the information exchanges are done through calls to classes and their methods. RabbitMQ, InfluxDB, Grafana and OpenStack can be installed on different computers using the HTTP RESTful APIs. Different protocols can be used to manage alarms. We chose Webhook, an HTTP callback that detects changes in the working conditions of a program.

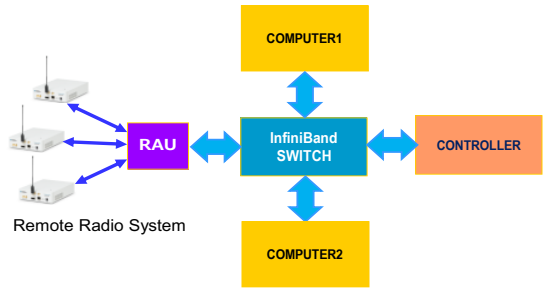

Fig. 3. OOCRAN testbed architecture.

*B. OOCRAN Testbed*

The testbed design pursues two objectives: (1) to enable the creation of tailored NSs from virtualized resources and (2) to create a working environment for dealing with RF signal generation, acquisition and processing in a cloud computing environment. OOCRAN allows running complex NSs in simulated channel mode, using IP connections between VNFs where the channel is represented by a VNF, and in radiated over-the-air mode, where the signal is transmitted by one antenna and received by another. The same set of VNFs can be used in both cases for processing the data flow [13].

The OOCRAN testbed is illustrated in Fig. 3 and features the following components:

- A **computing cluster** built around two SuperServer 6018TR-TF and one Intel i7 PC acting as the controller and administrating the C-RAN.
- **Infiniband switch** IS5022 with 640 Gb/s bandwidth and 100 ns port-to-port latency.
- A **remote radio system** (RRS) encompassing an Intel i7 PC, which acts as the radio aggregation unit (RAU), and several Ettus Universal Software Radio Peripherals (USRPs) acting as the remote radio units (RRUs). Additional RRUs—LimeSDR + ODROID single board computer—are being integrated.
- **VNFs** that allow building NSs using the srsLTE waveform library [14]. Several VMs that run different configurations of the LTE PHY layer for the LTE base station (eNB) and user equipment (UE) are available as VNFs.

## III. EXPERIMENTS

We implemented an LTE radio access network, featuring different eNodeB configurations to illustrate the capabilities of OOCRAN for deploying tailored NSs. OOCRAN monitors the computing and radio resources that a NS uses. It can instantiate NS reconfiguration. Here we demonstrate dynamic NS deployment.

*A. Scenario*

We consider the transition from a lightly loaded to a medium loaded cell. This corresponds to a typical fluctuation that a cellular operator experiences today, where some cells are more loaded than others and the load in each cell varies over the course of a day as well as before, during and after a special event. The lightly loaded cell requires only few resources and, thus, a simpler NS. The medium loaded cell requires more resources and another NS to satisfy the higher service demand. We leverage the system flexibility of LTE, which in its base configuration offers system bandwidths of 1.4, 3, 5, 10, 15, and 20 MHz.

Our case considers the 1.4 and 5 MHz LTE eNodeB for the two NSs. The 1.4 MHz system corresponds to the practical scenario of low network load, where the few UEs and their service requests can be met using only 1.4 MHz of RF spectrum and a relatively low computing load. The 5 MHz system needs about 3.5x the spectrum and, correspondingly, processes more samples and uses more computing resources.

OOCRAN is the MANO framework instance that manages the wireless infrastructure. The configuration and management of the VNFs is handled by the NS_ENGINE. The RF_ENGINE establishes the connection between the VNF and the RF front end and chooses the carrier frequency. The NS_ENGINE and RF_ENGINE thus constitute the VNF managers. MONITOR is waiting for the trigger that will change the NS. Therefore, MONITOR and GUI compose the orchestrator.

*B. Results*

Fig. 4 shows the baseline configuration, where the computing cluster executes OOCRAN on a VM, but no NS. The VM correspondingly shows low CPU and RAM usage.

The WSP deploys a simple NS to satisfy the user demands corresponding to a lightly loaded cell. During the NS operation, different UEs are served using some percentage of the available computing and radio resources. The WSP may need to pay for the maximum resource consumption corresponding to the NS; i.e. 1.4 MHz of spectrum and the computing resources to run the VNFs at full radio/user capacity. When the user demand increases and surpasses a threshold, the WSP may decide to reconfigure the NS to be able to provide more capacity and be able to effectively serve the expected increase in demand. The WSP therefore initiates a NS reconfiguration, where the new NS is built from other VNFs, VNF configurations or instances, and triggers a RAN reconfiguration from 1.4 to 5 MHz LTE.

Fig. 5 shows the NS reconfiguration which happens around 17:44 and takes a fraction of a minute. The old NS is stopped and its resources are freed and the new NS is started. This has an impact on CPU usage, which shows a short dip at around 17:44 and then increases to 15-20%. The RAM increases to 3.6 GB for the 5 MHz mode. The block error rate (BLER) and signal-to-noise ratio (SNR) remain excellent, as they before since the channel has not changed. The occupied resource blocks (RBs) increase from 6 to 25, corresponding to the 1.4 and 5 MHz system bandwidth (1 RB covers 12 subcarriers, where two adjacent subcarriers are separated by 15 kHz in LTE). Fig. 5 clearly indicates the moment of reconfiguration and shows the expected behavior in terms of resources.

## IV. CONCLUSIONS

This paper has presented our OOCRAN software framework and testbed and demonstrated it for NS creation, deletion and reconfiguration. OOCRAN enables rapid prototyping and testing with simulated channels, enabling reproducible experiments, and real channels, for demonstrating real-time signal processing and NS reconfiguration in a C-RAN.

Future work will analyze different ways of NS reconfiguration, such as stop and redeploy versus changing configuration parameters during runtime. We will consider different NS granularities and analyze optimal deployment strategies that maximize resource and network efficiencies at reasonable control and reconfiguration overhead. The reconfiguration time is an important metric that needs to be quantified to characterize the delay and the amount of data that need to be buffered when switching from one NS to another when there are active users. Operators can also use OOCRAN's support for multiple alarms to, for example, force the handover of active users before the NS reconfiguration is triggered; or they can implement other solutions to meet the high quality of experience expectations. Additional metrics need to be established and evaluated to facilitate smooth NS reconfiguration and end user transparency.

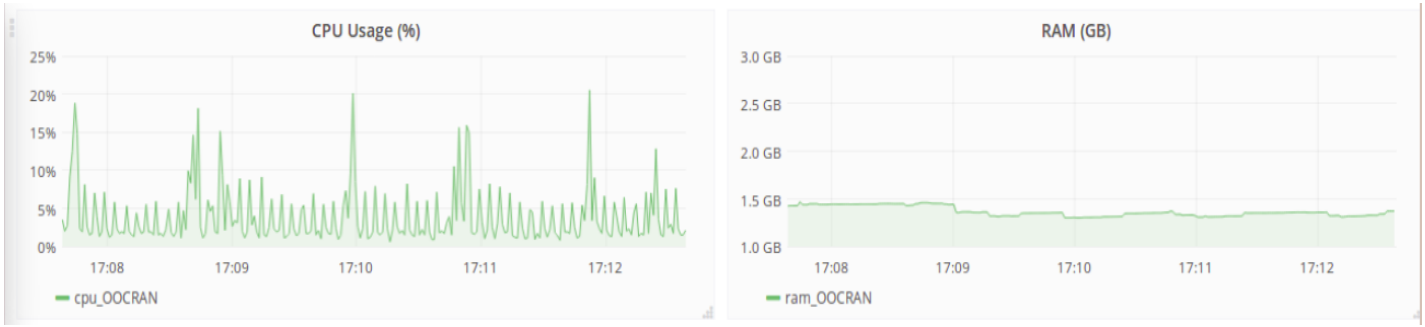

Fig. 4. Baseline CPU and memory usage when executing OOCRAN on a VM without a NS.

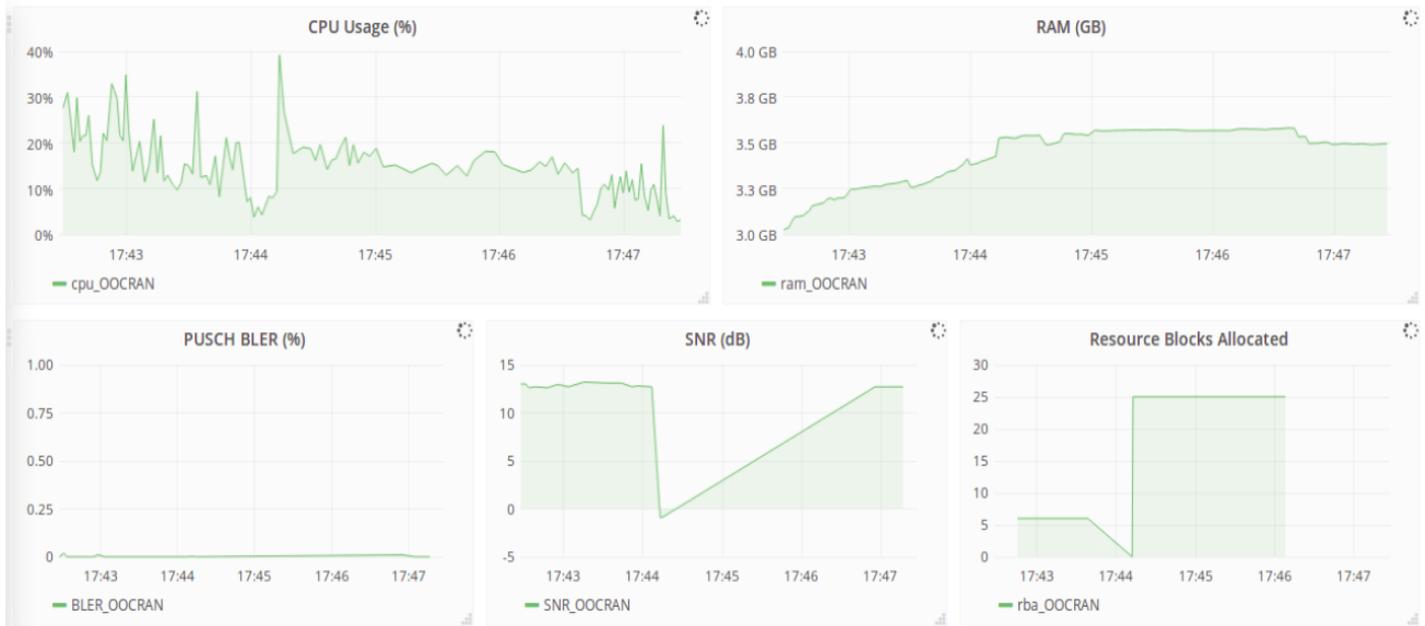

Fig. 5. NS reconfiguration from 1.4 to 5 MHz LTE at around the 17:44 time mark.